\documentstyle[12pt]{article}
\def\gsim{\mathrel{\rlap {\raise.5ex\hbox{$ > $}}
{\lower.5ex\hbox{$\sim$}}}}
\def\lsim{\mathrel{\rlap {\raise.5ex\hbox{$ < $}}
{\lower.5ex\hbox{$\sim$}}}}

\newcommand{\pr}{\paragraph{}}
\newcommand{\be}{\begin{equation}}
\newcommand{\ee}{\end{equation}}
\newcommand{\bea}{\begin{eqnarray}}

\newcommand{\eea}{\end{eqnarray}}

\newcommand{\nk}{\noindent}
\def\gappeq{\mathrel{\rlap {\raise.5ex\hbox{$>$}}
{\lower.5ex\hbox{$\sim$}}}}
 
\def\lappeq{\mathrel{\rlap{\raise.5ex\hbox{$<$}}
{\lower.5ex\hbox{$\sim$}}}}
 
\begin{document}
 
\begin{titlepage}
\begin{flushright}
ACT-12/97 \\
CTP-TAMU-33/97 \\
OUTP--97-37P \\
quant-ph/9708003 \\
\end{flushright}

\begin{centering}
\vspace{.1in}
{\large {\bf On Quantum Mechanical Aspects of 
Microtubules}} \\
\vspace{.2in}
{\bf N.E. Mavromatos$^{a}$} 
and 
{\bf D.V. Nanopoulos$^{b,c,d}$}

\vspace{.03in}
 
{\bf Abstract} \\
\vspace{.1in}
\end{centering}
{\small 
We discuss possible quantum mechanical aspects
of MicroTubules (MT), 
based on recent developments 
in quantum physics.
We focus  
on potential mechanisms for `energy-loss-free' transport 
along the microtubules, which could be considered
as realizations of Fr\"ohlich's ideas
on the r\^ole of solitons for superconductivity 
and/or biological matter. In particular, 
by representing the MT arrangements as {\it cavities},
we present a novel 
scenario on  
the formation of
macroscopic (or mesoscopic) 
quantum-coherent states, 
as a result of the (quantum-electromagnetic) 
interactions 
of the MT dimers with 
the surrounding 
molecules of the ordered water in the interior of the MT cylinders. 
Such states decohere 
due to dissipation through the walls of the MT. Transfer of 
energy without dissipation, due to such coherent modes, 
could occur only 
if the decoherence 
time is larger than the average time scale required for 
energy transfer across the cells.
We present 
some generic order of magnitude 
estimates of the decoherence time in a typical model for MT
dynamics. 
Our conclusion is that 
the quantum coherent states play a r\^ole in 
energy transfer  if the 
dissipation through the walls
of the MT cavities is fairly suppressed, 
corresponding to damping time scales $T_r \ge 10^{-4}-10^{-5}$ sec, 
for moderately large MT networks.  
We suggest specific experiments to test 
the above-conjectured quantum nature of 
the microtubular arrangements inside the cell. These experiments 
are similar in nature to those in atomic physics, used in 
the detection of the Rabi-Vacuum coupling between 
coherent cavity modes and atoms. Our conjecture is that a similar 
Rabi-Vacuum-splitting phenomenon occurs in the absorption (or emission) 
spectra of the MT dimers, which would constitute a 
manifestation of the dimer coupling with 
the coherent modes in the ordered-water environment (dipole quanta),
which emerge
due to `super-radiance'. }

\vspace{0.15in}
\begin{flushleft}
$^{a}$ P.P.A.R.C. Advanced Fellow, Department of Physics
(Theoretical Physics), University of Oxford, 1 Keble Road,
Oxford OX1 3NP, U.K.  \\
$^{b}$ Department of Physics, 
Texas A \& M University, College Station, TX 77843-4242, USA, \\
$^{c}$ Astroparticle Physics Group, Houston
Advanced Research Center (HARC), The Mitchell Campus,
Woodlands, TX 77381, USA, \\
$^{d}$ Academy of Athens, Chair of Theoretical Physics, 
Division of Natural Sciences, 28 Panepistimiou Avenue, 
Athens 10679, Greece. \\

\end{flushleft}

\end{titlepage} 

\newpage
\section{Introduction}

{\it MicroTubules} (MT) appear to be 
one of the most fundamental structures of the 
interior of living cells~\cite{hameroff}.
These are paracrystalline cytoskeletal
structures which seem to play a fundamental r\^ole 
for the cell {\it mitosis}.
It is also believed
that they play an important r\^ole 
for the transfer of electric signals
and, more general, of energy 
in the cell. 

In this latter respect it should be mentioned 
that energy transfer across the cells,  
without dissipation,  
had been conjectured to occur in 
biological matter 
by 
Fr\"ohlich~\cite{Frohlich}, already some time ago. 
The phenomenon conjectured by Fr\"ohlich was based 
on his one-dimensional superconductivity model:
in one dimensional electron systems with holes, 
the formation 
of solitonic structures due to electron-hole pairing
results 
in the transfer of electric 
current without dissipation.
In a similar manner, Fr\"ohlich conjectured 
that energy in biological matter could be transfered 
without dissipation, if appropriate solitonic 
structures are formed inside the cells.
This idea has lead theorists to construct 
various models for the energy transfer 
across the cell, based on the formation of kink 
classical solutions~\cite{lal}. 

In the early works no specific microscopic models had been 
considered~\cite{lal}.
Recently, however, after the identification of 
the MT as one of the most important structures of the cell,
both functionally and structurally, 
a model for their dynamics 
has been presented in ref. \cite{mtmodel}, 
in which the formation of solitonic structures,  
and their r\^ole in energy transfer across the MT, is discussed
in terms of classical physics.   
In ref. \cite{mn} we have considered the {\it quantum aspects}
of this one-dimensional model, and argued on the consistent 
quantization of the soliton solutions, as well as the fact that 
such semiclassical solutions may emerge as a result of `decoherence'
due to environmental entanglement, according to recent 
ideas~\cite{zurek}.  

The basic assumption of the model used in ref. \cite{mn}
was that the fundamental structures
in the MT (more specifically of the brain MT) 
are Ising spin chains (one-space-dimensional structures).
The interaction of each chain with the neighboring
chains and the surrounding water environment 
had been mimicked by suitable potential terms in the 
one-dimensional Hamiltonian.
The model describing the dynamics of such one-dimensional 
sub-structures 
was the ferroelectric distortive spin chain
model of ref. \cite{mtmodel}.

The one-dimensional nature of these fundamental
building blocks, then, opens up the way for a mathematical
formulation of the chain as a {\it completely intergrable} 
field theory model, characterized by an infinite-number
of conservation laws. The latter are associated with 
global excitation modes, 
completely {\it delocalised} in the chain space-time.
This integrability structure 
proves sufficient
in providing a satisfactory solution
to memory coding and capacity~\cite{mn}. Such features
might turn out to be important
for a model of the brain as a {\it quantum computer}.

This formulation allowed the authors of ref. \cite{mn}
to employ the so-called Liouville (non-critical) string theory
approach to decoherence~\cite{emn},
which is a formalism to treat such completely integrable 
spin systems embedded in a `two-dimensional quantum gravity 
environment'. At this point we should stress that 
such an environment does not {\it necessarily} represent 
realistic space-time effects. As stressed in ref. \cite{mn}, 
an external stimulus may distort the neighboring space-time 
of the unpaired spins of the MT dimers, and such a distortion 
may be {\it effectively} described by a two-dimensional 
`metric' in the Liouville picture of MT~\cite{mn},
which should not be necessarily 
identified with a relatistic four dimensional 
space time 
metric~\footnote{It is however, possible, that such a distortion
involves a sufficient movement of mass so as to 
create a virtual microscopic singularity 
in the surrounding space time~\cite{HP,ideas,mn}; this could then 
cause
collapse of the macroscopic quantum state, leading 
to {\it conscious perception}.
Such daring assumptions/conjectures
have gain some support by the fact that in 
certain physical models of MT dynamics~\cite{mn}
the estimated time of collapse 
is of order ${\cal O}(1~{\rm sec})$, which is 
in excellent qualitative agreement 
with a plethora of experimental/observational
findings in Neurobiology. In the present work we shall not 
discuss these issues.}.

The coupling of the MT chain with its `environment'
may lead, according 
to the analysis of ref. \cite{mn},  
to the formation of {\it macroscopic quantum coherent 
solitonic} states. Then, the details of the environment
would play a crucial r\^ole in determining the 
time scales over which such coherent states could be 
sustained {\it before collapsing to classical ground states}. 

Such analyses have yield
important results in the past~\cite{ehns,emohn,zurek,gisin}, 
concerning the passage from the quantum to classical 
world. In particular, in ref. \cite{zurek} it was suggested
that quantum coherent states may appear as a result of
decoherence, depending on the nature of the environment~\cite{albrecht}. 
Such states are minimum entropy/uncertainty states, which propagate 
`almost classically' in time, in the sense that their shape 
is retained during evolution. 
Such `almost classical' states 
are termed 
`pointer states' by Zurek~\cite{zurek}. 
In view of the results of 
ref. \cite{gisin}, concerning
the localization of the wave function
in open (stochastic) quantum-mechanical systems, the emergence of 
pointer states, as a result of decoherence, 
may be interpreted as implying 
that the localization process of the state vector had stopped
at a stage where 
it is not complete, but is such that the resulting (minimum-entropy) 
state
is least susceptible to the effects of the environment. 

Not all environments admit such pointer states~\cite{albrecht}.
Here lies one of the advantages of 
viewing the MT system of chains
as a completely integrable Liouville-string theory.
The model possesses a pointer basis~\cite{aspects},
which in ref. \cite{mn} has been identified with the 
quantized (via squeezed-coherent states) 
soliton solutions of ref. \cite{mtmodel}.
Another advantage lies on the 
fact that energy is {\it conserved} on the average
in the model, despite the 
environmental entanglement. This is for purely 
stringy reasons pertaining to the 
Liouville approach. The reader may find details
in ref. \cite{emn}.
Physically this means that such a formalism
is capable of describing {\it energy-loss-free} transport 
in biological cells, thereby providing a realization
of Fr\"ohlich's ideas~\cite{Frohlich}. 

In this article we focus our attention 
on the {\it microscopic nature} 
of such coherent states, and try to understand their emergence 
by as much `conventional' physics arguments as possible. By 
`conventional' we mean scenaria based on 
{\it quantum electrodynamics},
which are the major interactions expected to dominate 
at the energy scales of the 
MT~\footnote{According to the above 
discussion, then, quantum gravity 
entanglement 
could be, if at all, important for the decoherence
of the coherent state formed as a result of 
the water-dimer coupling. However this would require 
extreme isolation of the MT system, so that any conventional 
decoherence due to ordinary environmental entanglement
is suppressed. Such isolated MT subsystems have been conjectured
to occur in certain parts of brain, in which case  
quantum gravity plays a r\^ole in consciousness by
inducing a collapse of the coherent-state  of the MT
dimers~\cite{mn,HP,ideas}. 
At present, we do not know whether this is feasible inside the brain.}.
We make an attempt to pick up possible sources of 
decoherence, consistent with the above-mentioned
energy-loss-free transfer scenario in biological cells, and 
estimate the associated time scales.

The structure of this article is a follows: in section 2 we review 
briefly the physical one-dimensional model describing MT dynamics,
and discuss briefly the mathematical background for obtaining 
solitonic structures. An important r\^ole for the existence of the 
solitonic structures is played by the ordered
water in the interior of the MT. The importance of the water 
in the interior of the MT has been emphasized by S. Hameroff,
in connection with information processing, 
already some time ago~\cite{guides,ordered}.
In our article we present a scenario for the formation 
of electric-dipole quantum coherent modes in the water 
of the MT, inspired by earlier 
suggestions on `laser-like' behaviour of the water~\cite{prep,delgiud,jibu}. 
Such coherent modes 
couple with the unpaired 
electrons of the MT dimers.

In this work we conjecture that such a 
coupling 
would result in the 
so-called 
Vacuum Field Rabi 
Splitting (VFRS)~\cite{rabi}, a sort of dynamical Stark effect occuring 
in the emission or absorption spectra of the dimers 
as a result of the vaccum quantum fluctuations. 
This phenomenon is characteristic of the behaviour of 
atoms inside quantum electrodynamical 
cavities, the r\^ole of which is played in our work 
by the MT cylindrical structures
themselves.

A review of this phenomenon in the atomic physics is 
given in section 3.  
The atomic Rabi Splitting  is a consequence  
of the coupling 
of the atoms with 
the coherent modes of electromagnetic radiation inside the cavity. 
It is  
considered by many as a `proof' of 
the quantum nature of the electromagnetic radiation~\cite{rabi}.
In our work, we consider this phenomenon, if true, 
as a manifestation of the quantum mechanical nature of the 
MT arrangement inside the cell.

This is discussed in section 4, where  
we present a scenario according to which 
the MT arrangements inside the cell act as fairly 
isolated {\it cavities}. The ordered-water
molecules in their interior, then, provides an environment,  
necessary to form coherent quantum modes 
(dipole quanta)~\cite{prep,delgiud,jibu},
whose coupling 
with the dimers results in VFRS in the respective absorption 
spectra. Due to dissipation from the cavity walls, there is 
decoherence of the combined dimer-cavity system coherent state. 
The resulting decoherence time scales are estimated,
for various sources of envirnoments. 
The main conclusion of our analysis 
is that in order for the quantum coherent states to play a r\^ole 
in dissipationless energy transfer, i.e.  
the decoherence time to be larger than the 
the scale 
required for energy transfer in MT - 
estimated to be of ${\cal O}(5 \times 10^{-7}~sec)$,
for a moderately long MT,
in the model of ref. \cite{mtmodel} -   
the damping time scales of the MT cavities must be 
$T_r \ge 10^{-4}-10^{-5} {\rm sec}$.

Finally, in section 5, we present our conclusions and 
suggest some epxeriments
to test the above-conjectured quantum mechanical 
origin of MT arrangements inside the cell. 
The experiments are inspired by the physics of 
Rydberg atoms in electromagnetic cavities~\cite{rabiexp,rabiexp2}.
   
\section{Review of the Physical Model for Microtubule Dynamics} 
\subsection{Classical Considerations}

MicroTubules (MT) are hollow cylinders 
comprised of an exterior surface
(of cross-section
diameter
$25~nm$)
with 13 arrays
(protofilaments)
of protein
dimers
called tubulines.
The interior of the cylinder
(of cross-section
diameter $14~nm$)
contains {\it ordered water} molecules,
which implies the existence
of an electric dipole moment and an electric field.
The arrangement of the dimers is such that, if one ignores
their size,
they resemble
triangular lattices on the MT surface. Each dimer
consists of two hydrophobic protein pockets, and
has an unpaired electron.
There are two possible positions
of the electron, called $\alpha$ and $\beta$ {\it conformations}. 
When the electron is
in the $\beta$-conformation there is a $29^o$ distortion
of the electric dipole moment as compared to the $\alpha $ conformation.

In standard models for the simulation of the MT dynamics,
the `physical' degree of freedom -
relevant for the description of the energy transfer -
is the projection of the electric dipole moment on the
longitudinal symmetry axis (x-axis) of the MT cylinder.
The $29^o$ distortion of the $\beta$-conformation
leads to a displacement $u_n$ along the $x$-axis,
which is thus the relevant physical degree of freedom.
This way, the effective system is one-dimensional (spatial),
and one has the possibility of a quantum integrable
system~\cite{mn}. 

Information processing
occurs via interactions among the MT protofilament chains.
The system may be considered as similar to a model of
interacting Ising chains on a trinagular lattice, the latter being
defined on the plane stemming from fileting open and flatening
the cylindrical surface of MT.
Classically, the various dimers can occur in either $\alpha$
or $\beta$ conformations. Each dimer is influenced by the neighboring
dimers resulting in the possibility of a transition. This is
the basis for classical information processing, which constitutes
the picture of a (classical) cellular automatum.

The {\it quantum computer character} of the MT network results
from the {\it assumption} that each dimer finds itself in a
superposition of $\alpha$ and $\beta$ conformations \cite{HP}.
Viewed as a {\it two-state} quantum mechanical 
system, the MT tubulin dimers couple to conformational changes with
$10^{-9}-10^{-11} {\rm sec}$ transitions, corresponding to an
angular frequency $     \omega \sim{\cal O}( 10^{10})
-{\cal O}(10^{12})~{\rm Hz}$. In the present work we 
assume the upper bound of this frequency range 
to represent (in order of magnitude) the characteristic frequency 
of the dimers, viewed as a two-state quantum-mechanical system: 
\be
     \omega _0 \sim {\cal O}(10^{12})~{\rm Hz} 
\label{frequency2}
\ee

The scenario for quantum computation in MT presuposes that 
there exists a macroscopic
coherent state among the various chains. Let us now try to understand 
its emergence.
Let $u_n$ be the displacement field of the $n$-th dimer in a MT
chain.
The continuous approximation proves sufficient for the study of
phenomena associated with energy transfer in biological cells,
and this implies that one can make the replacement
\be
  u_n \rightarrow u(x,t)
\label{three}
\ee
with $x$ a spatial coordinate along the longitudinal
symmetry axis of the MT. There is a time variable $t$
due to
fluctuations of the displacements $u(x)$ as a result of the
dipole oscillations in the dimers.
At this stage, $t$ is viewed as a reversible variable.
\pr

The effects of the neighboring
dimers (including neighboring chains)
can be phenomenologically accounted for by an effective
potential $V(u)$. In the model of ref. \cite{mtmodel} 
a double-well potential was used, leading to a 
classical kink solution for the $u(x,t)$ 
field. More complicated interactions are allowed
in the string picture, as explained in ref. \cite{mn}. 
More generic polynomial potential have also been considered
in ref. \cite{mn}. 

The effects of the surrounding water molecules can be
summarized by a viscuous force term that damps out the
dimer oscillations,
\be
 F=-\gamma \partial _t u
\label{six}
\ee
with $\gamma$ determined phenomenologically at this stage.
This friction should be viewed as an environmental effect, which
however does not lead to energy dissipation, as a result of the
non-trivial
solitonic structure of the
ground-state
and the non-zero constant
force due to the electric field.
This is a well known result, directly relevant to
energy transfer in biological systems \cite{lal}.

In mathematical terms the effective equation of motion 
for the relevant field degree of freedom $u(x,t)$ reads:
\be
u''(\xi) + \rho u'(\xi) = P(u) 
\label{generalsol}
\ee
where $\xi=x-vt$, $v$ is the velocity of the soliton, 
$\rho \propto \gamma$~\cite{mtmodel}, 
and 
$P(u)$ is a polynomial in $u$, of a certain degree, stemming 
from the variations of the potential $V(u)$ describing interactions
among the MT chains~\cite{mn}.
In the mathematical literature~\cite{otinowski} 
there has been a classification of solutions of equations
of this form. For certain forms of the potential, the 
solutions
include {\it  kink solitons} that may be 
responsible for dissipation-free energy transfer in biological
cells~\footnote{In 
the `string picture of MT' the various solutions are related~\cite{mn}
by appropriate Renormalization-Scheme changes on the world-sheet
of the string (generalizaing appropriately the particle like structures). 
This is reflected in certain {\it ambiguities} in the form of the 
string potential 
$V(u)$; what seems to be a unique phenomenological potential 
in point-like theory, determining various forms of soliton 
solutions, may not have such a unique interpretation  in the 
completely-integrable (non-critical) string picture of MT. This 
suggests that the point-like field theory soliton
solutions, discussed above, 
constitute only a `low-energy' approximation of more complicated
(solitonic) ground states of the string.
As discussed in the context of Liouville
strings in ref. \cite{mn}, such {\it stringy ground states} 
share similar properties
with their point--like theory counterparts as far as dissipation-free
energy transfer is concerned.}. 
A typical propagation velocity of the kink solitons of ref. \cite{mtmodel}
is $v \sim 2~{\rm m/sec}$. 
This implies that 
for moderately long microtubules, 
of length $L \sim 10^{-6}$ m, such kinks transport energy in 
\be
      t_F \sim 5 \times 10^{-7}~{\rm sec}
\label{FS}
\ee
This scale is larger than the time scale 
that Fr\"ohlich had conjectured as corresponding 
to the frequency of the coherent 
phonons in biological matter ($t \sim 10^{-11}-10^{-12}$ $s$).
In this article, however, we shall keep calling the time (\ref{FS}) 
the Fr\"ohlich scale, since, upon quantization, the kink soliton 
solutions of ref. \cite{mtmodel} yield  quantum coherent
states~\cite{mn} that are very similar to Fr\"ohlich's phonons.

Therefore, in the above `phenomenological' approach to the
MT physics, 
the importance of the water environment can be seen 
formally as follows: were it not for the friction term (\ref{six}) 
there would be no stable solitonic structures in the 
ferroelectric distortive model of ref. \cite{mtmodel}~\footnote{In the 
(non-critical) Liouville 
string picture such a coupling 
is responsible for the appearance 
of the appropriate `dilaton'-like terms in the world-sheet 
action, which are essential for consistency 
of the Liouville string~\cite{DDK,aben,mn}.}. 

\subsection{Quantum Considerations}

Let us attempt a microscopic 
analysis of the physics underlying the interaction
of the water molecules with the dimers of the MT. 
Our investigation points towards the fact that, as a result of 
the ordered structure of the water environment,
there appear {\it collective} coherent modes,
which in turn interact with the dimer structures (mainly through 
the unpaired electrons of the dimers)
leading to the formation of a quantum coherent 
solitonic state that extends over the whole network of MT. 
According to the idea put forward in ref. \cite{mn},
following ref. \cite{zurek}, such coherent states
should be viewed as the result of {\it decoherence} of the 
dimer system due to its interaction/coupling with the 
water environment. 

As we shall argue below, such a coupling could be detected 
by 
a phenomenon analogous to what is happening 
in atoms interacting with coherent modes 
of the electromagnetic radiation in 
{\it Cavities}, namely the
{\it Vacuum-Field Rabi Slitting} (VFRS)~\cite{rabi}. 
Our {\it conjecture} is that 
the interior of MT, full of ordered water 
molecules,  can be viewed as a {\it cavity}
rather than a {\it wave guide}; the cavity 
structure can be formed by `closing'
the ends of the MT, and is a way 
of providing a fairly isolated system
which can sustain coherent modes. 

This type of environmental entanglement
is described in the Liouville framework
by
the
world-sheet renormalization-group $\beta$-function, $\beta^u$, 
of the displacement field, 
$u(x)$, of the dimers, decribed above.
A non-trivial $\beta^u \ne 0$ expresses 
deviations from {\it conformal invariance}, as a consequence  
of environmental entanglement~\cite{mn}. 
The $\beta$-function corresponds to an energy-dependent environmental coupling,
since its magnitude depends on the energy 
of the field $u(x)$~\cite{mn}:
\be
   \beta^u = {\cal O}[(E/M_s)^2]
\label{betau}
\ee
where $M_s$ is a string scale, characteristic of the problem.
The scale $M_s$ acts as an ultraviolet 
cut-off for the energies of the low-energy field theory
obtained from the `string', and its size depends on the details of 
the subsystem $u(x,t)$ as well as the 
environment.

The Master Equation 
for the time evolution 
of the density matrix $\rho $ of the displacement field $u$ in the Liouville 
approach reads~\cite{emn,mn}:
\be
    \partial_t \rho =  i[\rho, H] + i\beta^u[\rho, u]M_s
\label{displacementmaster}
\ee
where $H$ is the Hamiltonian of the dimer chain.
The decoherence time in this approach is estimated as~\cite{mn}:
\be
     t_{decoh/Liouv} \sim \frac{1}{{\cal N}\beta^u M_s}
\label{liouvdecoh}
\ee
where ${\cal N}$ denotes the number 
of dimers in the MT chain.
The time scale (\ref{liouvdecoh}) 
is the time  over which the solitonic quantum coherent pointer
states are formed in the MT networks, as a result of the 
ordered-water-induced decoherence. 

We now remark that, in the approach
of viewing the MT as cavities, 
the main source of dissipation will then be 
the leakage of photons or other coherent modes
from the cavity. 
Its rate  can be assumed {\it small}
for the MT of the brain, otherwise the incoherent 
mode will dominate. 
This dissipation constitutes an ordinary environment,
which cannot lead to the formation of pointer states, but 
instead 
induces eventual collapse of 
the solitonic states into completely classical 
ground states. 

Hence, according to the above 
discussion, there are two stages when decoherence plays an 
important r\^ole. 
\begin{itemize}
\item{(i)} The first stage concerns the coupling of the 
system of dimers with the coherent modes 
formed in the ordered water. This coupling 
pertains to a specific environment which induces
decoherence in the dimer sub-system.
producing 
quantum coherent states (`pointer states') according 
to refs. \cite{zurek,mn}. 
The time scale of this decoherence
corresponds to the time scale necessary for the formation 
of the solitonic coherent states of ref. \cite{mn},
see eq. (\ref{liouvdecoh}) above.

\item{(ii)} The second stage refers to the decoherence due to 
ordinary dissipation through the walls of the 
(imperfect) MT cavities. Such an ordinary environment
does not admit a pointer basis~\cite{albrecht}, but causes
collapse of the quantum coherent state of stage (i) 
down to a classical ground state. 

\end{itemize}

In this article, 
we shall estimate the decoherence time scales
of both stages (i) and (ii) and compare them with the 
Fr\"ohlich scale. Moreover, we shall propose specific experimental tests
for the detection of the quantum-mechanical behaviour of the MT arrangements
inside the cell. An important feature of all these tests is,
as we mentioned previously,  
the Vacuum-Field Rabi Splitting phenomenon, 
to a brief description of 
which 
we now turn for instructive purposes.

\section{On the Rabi Splitting in Atomic Physics and the Quantum Nature of 
Electromagnetic Radiation} 

\subsection{Description of the Rabi splitting phenomenon}

In this section we shall recapitulate briefly
the VFRS in atomic physics. 
This phenomenon has been 
predicted for the emission spectra of 
atoms inside electromagnetic cavities~\cite{rabi}, in an attempt
to understand the {\it quantized nature} of the electromagnetic
radiation.   
The basic principle underlying the phenomenon is 
that, in the presence of an interaction among two oscillators
in resonance, 
the frequency degeneracy is {\it removed} by an amount proportional 
to the strength of the coupling. In the cavity $QED$ case 
of ref. \cite{rabi}, one oscillator consists of 
a small collection of $N$ atoms, whilst the other is 
a resonant mode of a high-$Q$(uality) cavity~\footnote{The quantity $Q$ is 
defined as the ratio of the 
stored-energy to the energy-loss per period~\cite{haroche},
by making the analogy with a damped harmonic oscillator.}.  
Immediately after the suggestion of ref. \cite{rabi},
a similar phenomenon has been predicted for absorption spectra
of atoms in cavities~\cite{agar}. 

By now, the situation has been verified experimentally
on a number of occasions~\cite{rabiexp}. In such experiments
one excites the coupled atom-cavity 
system by a tuneable field probe. The excitation is then found resonant 
{\it not} at  the `bare' atom or cavity frequencies but 
at the {\it split} frequencies of the `dressed' atom-field system. 
The spliting is enhanced for collections of atoms.
For instance, as we shall review below, 
for a system of $N$ atoms, the split is predicted to be~\cite{agar}:
\be
     {\rm Rabi~splitting}=2\lambda \sqrt{N} 
\qquad 2\lambda={\rm Rabi~splitting~of~a~single~atom}
\label{rabienhanced}
\ee
Despite its theoretical prediction by means of 
quantum mechanical oscillator systems coupled with 
a {\it quantized} radiation field mode in a cavity, 
at present there seems
to be still a {\it debate} 
on the nature of the phenomenon: (i) the dominant opinion
is that the Rabi splitting is a manifestation of the {\it quantum
nature} of the electromagnetic radiation (cavity field), and is 
caused as a result of an {\it entanglement} between the atom and the cavity 
coherent modes of radiation. It is a sort of Stark effect, 
but here it occurs in the 
absence of an external field~\cite{rabi}. This `dynamical Stark effect' 
is responsible for a splitting of the resonant lines 
of the atoms by an amount proportional to the collective atomic-dipole 
amplitude. 
(ii) there is however a dual 
interpretation~\cite{dual}, which claims 
that the splitting can be observed in optical cavities as well,
and it is simply a result of {\it classical} wave mechanics inside the 
cavity, where the atomic sample behaves as a {\it refractive medium} 
with a {\it complex} 
index, which splits the cavity mode into two components.

Irrespectively of this second classical interpretation, one {\it cannot deny}
the presence of the phenomenon 
in entangled atom-quantum-coherent mode systems.
This is the point of view we shall be taking in this work, 
in connection with our picture of viewing MT filled with ordered water
as cavities. We shall try to make specific experimental
predictions that could shed light in the formation 
of quantum coherent states, and their eventual decoherence.
As mentioned above, the latter  
could be due to the interaction of the dimer unpaired spins
(playing the r\^ole of the atoms in the Rabi experiments) 
with the ordered-water coherent modes (playing the r\^ole 
of the cavity fields). Possible scenaria for the origin of such cavity
coherent modes will be described below. 

Let us first recapitulate briefly the theoretical 
basis of the Rabi-splitting phenomenon, which will allow the non-expert reader
to assess the situation better. We shall present 
the phenomenon from a point of view that will help us 
transcribe it directly to the MT case. 
Consider an atom of a frequency $\omega_0$ 
in interaction with a single coherent
mode of electromagnetic radiation field of frequency $\omega$.
The relevant 
Hamiltonian is:
\be
H=\hbar\omega_0\sum_{i} S_i^z + \hbar \omega a^\dagger a 
+ \sum_{i} (\hbar \lambda S_i^+ a + H.C.)
\label{hamrabi}
\ee
where $a^\dagger,a$ are the creation and annihilation 
cavity radiation field modes, 
$S_i^z$,$S_i^{\pm}$ are the usual spin-$\frac{1}{2}$ 
operators, and $\lambda$ is the atom-field coupling. 
The atom-field system is not an isolated system, since there is 
{\it dissipation} due to the interaction of the system with the 
surrounding world. An important source of dissipation is 
the leakage of photons from the cavity at some rate $\kappa$.
If the rate of dissipation is not too big, then 
a quantum coherent state can be formed, which would allow 
the observation of the vacuum-field Rabi oscillations. 
The density matrix $\rho$ of the atom-field system obeys a Markov-type 
master equation for the evolution 
in time $t$~\cite{agar}:
\be
\partial_t \rho =-\frac{i}{\hbar} [H, \rho] - \kappa (a^\dagger
a \rho - 2 a \rho a^\dagger + \rho a^\dagger a )
\label{markovrabi}
\ee
The limit $\kappa << \lambda \sqrt{N}$ guarantees
the possibility of the formation of a quantum coherent 
state, i.e. this limit describes environments that 
are weakly coupled to the system, and therefore the 
decoherence times (see below) are very long. 
In this limit one can concentrate on the off-diagonal elements
of the density matrix, and make the following 
(`secular') approximation for their evolution~\cite{agar}:
\be
\partial _t \rho _{ij} =-\frac{i}{\hbar} 
(E_{i} - E_{j} )\rho_{ij} - \Gamma _{ij} \rho_{ij} 
\label{secular}
\ee
where $\Gamma _{ij}$ denotes the damping factor, pertaining 
to the weak coupling of the atom-field system 
to the environment.  The analysis of ref. \cite{agar}
pertained to the evaluation of the susceptibility tensor
of the system, $\chi_{\alpha\beta}$,
which can be calculated by considering 
its interaction of the system with 
an external field of frequency $\Omega$. 
The absorption spectrum is proportional to 
${\rm Im}\chi (\Omega) $. A standard quantum-mechanical 
computation 
yields the result~\cite{agar}:
\bea
 &~&{\rm Im}\chi(\Omega) ={\rm cos}^2\theta
\frac{\Gamma _-/\pi}{\Gamma _-^2 + 
\{ \Omega - \omega_0 + \Delta/2 -\frac{1}{2}
(\Delta ^2 + 4 N \lambda ^2 )^{1/2}\}^2}
+ \nonumber \\
&~&{\rm sin}^2\theta \frac{\Gamma _+/\pi}{\Gamma _+^2 + 
\{ \Omega - \omega_0 + \Delta/2 + \frac{1}{2}
(\Delta ^2 + 4 N \lambda ^2 )^{1/2}\}^2}
\label{suscept}
\eea
with $\Delta \equiv \omega_0 - \omega$.
In the above expression the damping factors $\Gamma _{\pm}$ 
represent the damping in the equation of motion for the 
off-diagonal element of the density matrix $<\Psi_0|\rho|\Psi_{\pm}^{S,C}>$
where $\Psi_{\pm}$ are eigenfucntions of $H$, classified by the eigenvalues 
of the operators $S^2$, and $S^z+a^\dagger a \equiv C$, in 
particular~\cite{agar}:
$S=N/2$, and $C=1 -N/2$.

The expression (\ref{suscept}) summarizes the 
effect of Rabi-vacuum splitting in absorption spectra of atoms: 
there is a doublet structure (splitting) of the absorption spectrum
with peaks at:
\be
\Omega = \omega _0 - \Delta/2 \pm \frac{1}{2}( \Delta ^2 + 
4 N \lambda ^2 )^{1/2}
\label{rabiabs}
\ee
For resonant cavities the splitting occurs with equal weights
\be
  \Omega = \omega_0 \pm \lambda \sqrt{N} 
\label{rabisplitting}
\ee
Notice here the {\it enhancement} in the effect 
for multi-atom systems $N >> 1$. 
This is the `Vacuum Field Rabi Splitting phenomenon', predicted
in emission spectra in ref. \cite{rabi}. As we have already
mentioned, the above derivation
pertains to absorption spectra, where 
the situation is formally much simpler~\cite{agar}.
It is also this latter case that is of interest to us for the purposes
of this work. 

The quantity  $2\lambda \sqrt{N}$ is called the `Rabi frequency'~\cite{rabi}. 
{}From the emission-spectrum theoretical analysis 
an estimate of $\lambda$ may 
be inferrred which involves 
the matrix element, ${\underline d}$, of atomic electric dipole 
between the energy states
of the two-level atom~\cite{rabi}: 
\be 
   \lambda = \frac{E_{vac}{\underline d}.{\underline \epsilon}}{\hbar}
\label{dipolerabi}
\ee
where ${\underline \epsilon}$ is the cavity (radiation)  mode 
polarization, and 
\be
E_{vac} \sim  \left(\frac{2\pi \hbar \omega_c}{\varepsilon _0 V}\right)^{1/2} 
\label{amplitude}
\ee
is the r.m.s. vacuum field amplitude at the center 
of the cavity of volume $V$, and of frequency $\omega_c$,
with $\varepsilon _0$ the 
dielectric constant of the vacuum~\footnote{For cavities 
containing other dielectric media, e.g. water in the case of the MT,
$\varepsilon _0$ should be replaced by the dielectric constant 
$\varepsilon$ of the medium.}: $\varepsilon_0 c^2 =\frac{10^7}{4\pi}$,
in M.K.S. units. 
As mentioned above, 
there are simple experiments which confirmed this effect~\cite{rabiexp},
involving beams of Rydberg atoms resonantly 
coupled to superconducting cavities.

The situation which is of interest
to us
involves atoms that are {\it near resonance} with the cavity.
In this case $\Delta << \omega_0$, but such that 
$\lambda^2 N/|\Delta |^2 << 1$; in such a case, formula 
(\ref{rabiabs}) yields two peaks that are characterized
by dispersive frequency shifts $\propto \frac{1}{\Delta}$:
\be
    \Omega \simeq \omega_0 \pm  \frac{N\lambda ^2}{|\Delta|} 
+ {\cal O}(\Delta)
\label{dispersive}
\ee
whilst no energy exchange takes place between atom and cavity mode. 
This is also the case of interest in experiments 
using such Rabi couplings to construct  
Schr\"odinger's cats in the laboratory, 
i.e. macro(meso)scopic combinations
`measuring apparatus + atoms' to verify decoherence 
experimentally. The first experiment of this sort,
which confirms theoretical expectations, is described in ref.~\cite{brune}.
\pr
Another important issue, which has been used above, and 
in ref. \cite{brune},
is the {\it dephasing} of the atom as a result of the 
above-described atom-field Rabi entanglement. 
To understand better the situation, let us 
discuss a more generic case, that of a three-state
atom, $f,e,g$, with energies $E_g > E_e > E_f$. 
Suppose one is interested in the transition $f \rightarrow e$ 
by absorption, in the presence of atoms in interaction with a cavity
mode. Calling $D_{ef}^+ \equiv |e><f|$,$D_{ef}^- \equiv |f><e|=
(D_{ef}^+)^\dagger$, we have the effective Hamiltonian
for the transition $f \rightarrow e$~\cite{qnd}:
\be
 H_{eff}^{ef}=\hbar \omega_{eff}D_{ef}^+D_{ef}^-
\qquad ; \qquad \omega_{eff}=\omega_{ef} + \frac{\lambda^2 n}{\Delta}
\label{eftr}
\ee
where the `effective' frequency $\omega_{eff}$ incorporates 
the dispersive
frequency shifts (\ref{dispersive}) 
of the Rabi effect, appropriate for near-resonant
atom-cavity-field systems; $n$ is the number of cavity 
photons~\footnote{Notice that the $\sqrt{n}$ 
characteristic scaling law for the Rabi splitting
(\ref{rabisplitting})
is {\it also valid} in the case of interaction  
of a {\it single} atom with $n$ cavity oscillator quanta (e.g. a 
coherent cavity mode).}.

Consider now an experiment to measure, say, the 
photon number $n$ in the cavity. The relevant probe
$P$ can be the above-described three-state atom, in a superposition
of $e$ and $f$ states. In this picture the photon number $n$ is 
an eigenvalue of the cavity signal operator $a_s^\dagger a_s$;
the
interaction Hamiltonian between atom and cavity then reads~\cite{qnd}:
\be
H_I =\frac{\hbar\lambda^2}{\Delta} a_s^\dagger a_s D_{ef}^+D_{ef}^-
\label{iham}
\ee
The probe observable is the atomic dipole 
operator:
\be
   A_P=\frac{1}{2i}(D_{ef}^+-D_{ef}^-)
\label{dipole}
\ee
The Heisenberg evolution of this operator yields:
\be
i\hbar \frac{d}{dt}A_P = [A_P, H_{eff}^{ef}+H_I]
\label{evolution}
\ee
from which it is easily seen that in a time 
interval $t$ the phase of the probe is changing 
by:
\be
  \Delta \phi = \omega_{ef}t + \frac{\lambda^2 n}{\Delta}t
\label{phase}
\ee
The case of interest for the experiment of ref. \cite{brune}
is a two-state atom; the resulting phase shift
is then obtained from (\ref{phase}) by setting 
$\omega_{ef} =0$. In that case, 
the phase entanglement due to the atom-field Rabi coupling 
is
\be
   \Delta \phi _R = \frac{\lambda ^2 n}{\Delta}t
\label{rabiphase}
\ee
for a near resonance atom-field system, with a small 
detuning $\Delta $. Such a situation is encountered
precisely  in the experiment of ref. \cite{brune},
which will constitute our prototype for the extension 
of the above results to the MT case. 
Before doing so it is instructive to review 
briefly the experiment of \cite{brune}, where decoherence
of an open mesoscopic system (`Schr\"odinger's cat')
{\it has been demonstrated}. 

\subsection{Rabi splitting and decoherence: experimenal verification}

The experiment consists of sending a Rubidium atom
with two circular Rydberg states $e$ and $g$,
through a microwave cavity storing a small coherent field
$|\alpha>$. The coherent cavity mode is mesoscopic in the sense of
possessing 
an average number of photons of order ${\cal O}(10)$. 
The atom-cavity coupling is measured by the Rabi frequency
$2\lambda /2\pi=48~kHz$. The condition for Rabi dispersive
shifts (\ref{dispersive}) is satisfied 
by having a $\Delta/2\pi \in [70 , 800 ]~kHz$.

The set up of the experiment is as follows:  
The atom is prepared in the superposition of $e,g$ states, 
by virtue of a resonant microwave cavity $R_1$. 
Then it crosses the cavity $C$, which is coupled to a 
reservoir that damps its energy ({\it dissipation})
at a characteristic time scale $T_r << 1.5. ms$. 
Typical cavities, used for atomic scale experiments
such as the above, have 
dimensions which lie in the $mm$ range or at most 
$cm$ range.
In the cavity $C$ a number of photons varying 
from $0$ to $10$ is injected by a pulse source. 
The field in the cavity relaxes to vacuum - thereby 
causing {\it dissipation}
through leakeage of photons through the cavity -
during a time $T_r$, before being regenerated for the 
next atom. The experiment is at an effective 
temperature of $T=0.6K$, which is low enough so as to minimize
thermal effects. 
After leaving $C$, the atom passes through a second
cavity $R_2$, identical to $R_1$; one then measures the 
probability of finding the atom in the state, say, $g$. 
This would mean decoherence. 
The decoherence time is measured for various 
photon numbers; this helps testing the 
theoretical predictions that decoherence
between two `pointer states' of a quantum superposition
occurs at a rate proportional to the square of the distance
among the states~\cite{markov,milburn,zurek,emohn}. 

Let us understand this latter point better. 
The coherent oscillator states, characterizing 
the cavity modes, constitute such a pointer basis:
an oscillator in a coherent state is 
defined by the average number of oscillator quanta $n$
as: $|\alpha >:~|\alpha|=\sqrt{n}$. 
Then, consider the measurement of the above-described 
experiment, according to which there is only a phase entanglement
between the cavity and the atom: the combined 
atom-cavity (meter) system is originally in the state
\be
   |\Psi >=|e,\alpha e^{i\phi}>+ |g, \alpha e^{-i\phi}> 
\label{super}
\ee
According to (\ref{rabiphase})
the dephasing is atomic-level dependent
$\phi \propto \lambda^2 t n/\Delta$. 
Coupling the oscillator to a reservoir, that damps its energy 
in a characteristic time scale $T_r$, produces decoherence.
According to the general theory~\cite{ehns,milburn,qnd,markov}
the latter 
occurs during a time scale inversely proportional to 
the square of the distance between the `pointer' states $D^2$:
\be
   t_{collapse} = \frac{2T_r}{D^2} 
\label{decoh}
\ee
This is easy to verify in a prototype toy model, 
involving only a phase entanglement, i.e. without 
energy dissipation. To this end, consider the Hamiltonian:
\be
   H=\hbar\omega a^\dagger a + a^\dagger a \Gamma 
\label{phasedamp}
\ee
with $a^\dagger,a$
creation and annihilation operators of a quantum oscillator, and 
$\Gamma $ a phase damping. The `pointer' states
for this problem are characterized by the eigenvalues $n$ 
of the number operator $a^\dagger a$ which commutes
with the Hamiltonian $[ a^\dagger a~,~H]=0$, i.e. 
the pointer basis is $\{ | n > \}$. 
The pertinent Markov master equation
for the density matrix, $\rho $, reads: 
\be
\partial_t \rho =\frac{\kappa}{2}(2a^\dagger a \rho a^\dagger a - 
\rho a^\dagger a a^\dagger a - a^\dagger a a^\dagger a \rho ) 
\label{masterphasedamp}
\ee
Writing $\rho(t)=\sum_{m,n} \rho_{mn}(t)|n><m|$ it is straightforward
to determine the time dependence of $\rho_{nm}(t)$ 
from (\ref{masterphasedamp}) 
\be
   \rho_{nm}(t) = e^{-\kappa (n-m)^2t/2}\rho_{nm}(0)
\label{decoherencedamped}
\ee
which implies that the coherence  
between two different ($m \ne n$) pointer (number) states
is damped at a typical scale given by (\ref{decoh}),
where $T_r = \frac{1}{\kappa}$, and the distance $D$ 
is given by $n-m$ in the above example .  
As mentioned previously, this inverse $D^2$-behaviour 
seems to be a {\it generic} feature 
of open (decohering) systems, including open string models~\cite{emn,mn}.

In the set up of ref. \cite{brune} the distance $D$ is 
given by 
\be
   D=2\sqrt{n}{\rm sin}\phi \simeq 2n^{3/2}\frac{\lambda^2 t}{\Delta}   
\label{distance} 
\ee
for Rabi couplings $2\lambda$, such that $\lambda^2 t n << \Delta $. 
For mesoscopic systems $n \sim 10$, $D > 1$, and, hence, decoherence 
occurs over a much shorter time scale than $T_r$; in particular,
for $\Delta/2\pi \sim 70kHz$ the decoherence time 
is $0.24 T_r$~\cite{brune}. 
This concludes the construction of a Schr\"odinger's cat,
and the associated `measurement process'. 
Notice that the above construction is made in two stages:
first it involves an interaction of the atom with the 
cavity field, which results in a  coherent state of the 
combined `atom-meter', and then {\it dissipation} is induced 
by coupling the cavity (measuring apparatus) to the environment,
which damps its energy, thereby inducing {\it decoherence}
in the `atom+meter' system. The important point to realize is
that the more macroscopic the cavity mode is (i.e. the higher 
the number of oscillator quanta), the shorter the decoherence
time is. This is exactly what was to be expected 
from the general theory~\cite{zurek,ehns,emohn,emn}.

\section{Rabi Vacuum Coupling in MT and Energy Transfer in the Cell} 

\subsection{Microscopic Mechanisms for the formation of 
Coherent States in MT}

Above we have sketched the 
experimental construction of 
a mesoscopic quantum coherent state
(a `Schr\"odinger's cat' (SC)). 
The entanglement of the atom with the coherent 
cavity mode, manifested experimentally by the `vacuum Rabi splitiing',
leads to a quantum-coherent state for the combined atom-cavity 
system (SC), comprising of the 
superposition of the states of the two-level Rydberg atom.
Dissipation induced by the leakage of photons in the cavity 
leads to decoherence of the coherent atom-cavity state,
in a time scale given by (\ref{decoh}).  
This time scale depends crucially on the nature of the 
coupled system, and the nature of the `environment'. 
It is the point of this session to attempt to discuss a similar 
situation that conjecturally occurs 
in the systems of MT. We believe that understanding the formation of
Schr\'odinger's cats in MT networks will unravel, if true, 
the mysteries
of the brain as a (quantum) computer, which might also be related to
the important issue of `conscious perception', as advocated in refs. 
\cite{ideas,HP,mn}.

The first issue concerns the nature of the 
`cavity-field modes'. Our point in this section 
is to argue that 
the presence of {\it ordered water},
which seems to occupy the interior of the microtubules~\cite{ordered},
plays an important r\^ole in producing coherent 
modes, which resemble those of the ordinary electromagnetic 
field in superconducting cavities, discussed above. 

Let us first review briefly some suggestions
about the r\^ole of the electric dipole moment 
of water molecules in producing coherent modes after 
coupling with the electromagnetic radiation field~\cite{prep}. 
Such a coupling implies a `laser-like' behaviour. 
Although it is not clear to us whether such a behaviour characterizes
ordinary water, in fact we believe it does not due to the strong  
suppression of such couplings 
in the disordered ordinary water, however 
it is quite plausible that such a behaviour characterizes 
the ordered water molecules that exist in the interior of MT~\cite{ordered}. 
If true, then this electric dipole-quantum radiation
coupling will be responsible, according to the analysis of
ref. \cite{prep}, for the appearance of {\it collective}
quantum coherent modes. 
The Hamiltonian used in the theoretical model of ref. \cite{prep}
is:
\be
   H_{ow} = \sum_{j=1}^{M} [\frac{1}{2I} L_j^2 + {\underline 
A}.{\underline d}_{ej}]
\label{orwater}
\ee
where $A$ is the quantized
electromagnetic field in the radiation gauge~\cite{prep},
$M$ is the number of water molecules, $L_j$ is the
total angular momentum
vector of a single molecule, 
$I$ is the associated (average) moment of inertia,
and $d_{ej}$ is the  
electric dipole vector
of a single molecule, $|d_{ej}| \sim 2e \otimes d_e$,
with $d_e \sim 0.2$ Angstr\"om. 
As a result of the dipole-radiation interaction in 
(\ref{orwater}) coherent modes emerge, which in ref. \cite{delgiud}
have been interpreted as arising from the quantization of the
Goldstone modes responsible for the {\it spontaneous breaking}
of the electric dipole (rotational) symmetry. Such modes are termed
`dipole quanta' in ref. \cite{delgiud}. 

This kind of mechanism has been applied to microtubules~\cite{jibu},
with the conclusion that such coherent modes cause 
`super-radiance', i.e. create a specific quantum-mechanical 
ordering in the water molecules with characteristic times much shorter
than those of thermal interaction. In addition, the optical medium inside the 
internal hollow core of the microtubule
is made transparent by the coherent photons themselves~\cite{jibu}.
Such  
phenomena, if observed, could verify the coherent-mode emission 
from living matter, thereby verifying Fr\"ohlich's ideas.

In our picture of viewing the MT arrangements as cavities,
these coherent modes 
are the quantum coherent `oscillator' modes of section 3, 
represented by annihilation and creation operators 
$a_c,a^\dagger_c$, which play the r\^ole of the cavity modes, if 
the ordered-water interior of the MT is viewed as an isolated 
cavity~\footnote{This was not the picture envisaged in ref. \cite{prep}.
However, S. Hameroff, as early as 1974, had conjectured the 
r\^ole of MT as `dielectric waveguides' for photons~\cite{guides},
and in ref. \cite{jibu} some detailed mathematical construction 
of the emergence of coherent modes out of the ordered water are presented.
In our work in this article 
we consider the implications of such coherent modes
for the system of dimers, in particular for the formation of 
kink solitons of ref. \cite{mtmodel}. Thus, our approach is 
different from that in refs. \cite{prep,jibu}, where attention 
has been concentrated only on the properties of the water molecules.
We should emphasize that the phenomenon of optical 
transparency due to super-radiance may co-exist with the 
formation of kink soliton coherent states along the 
dimer chains, relevant to the dissipation-free energy 
transfer along the MT, discussed in the present work.}.
The r\^ole of the small collection of atoms, described in the 
atomic physics analogue above, is played in this picture by the 
protein dimers of the MT chains. The latter constitute a two-state system due to 
the $\alpha$ and $\beta$ conformations, defined by the position 
on the unpaired spin in the dimer pockets. 
 The presence of unpaired electrons 
is crucial to such an analogy. 
The interaction of the dipole-quanta coherent modes with the protein dimers
results in an entanglement which we claim is responsible for the 
emergence of {\it soliton quantum coherent states}, extending over 
large scale, e.g. the MT or even the entire MT network.

The issue, we are concerned with here, is whether 
such coherent states 
are 
responsible for {\it energy-loss-free
transport}, as well as for {\it quantum computations}
due to their eventual collapse, 
as a result of `environmental' entanglement
of the entire `MT dimers $+$ ordered water'
system. 
An explicit construction of such solitonic states
has been made in the field-theoretic model for MT dynamics of ref. \cite{mn}, 
based on classical ferroelectric models for the displacement 
field $u(x,t)$ discussed in section 2~\cite{mtmodel}. The quantum-mechanical
picture described here should be viewed as a simplification 
of the field-theoretic formalism, which, however, is sufficient 
for qualitative estimates of the induced decoherence.  

\subsection{Decoherence and dissipationless energy transfer}

To study quantitatively the effects of decoherence in MT 
systems we  
make the plaussible {\it assumption} that 
the environmental entanglement of the `ordered-water cavity' (OWC),
which is responsible for dissipation,
is 
attributed {\it entirely} to the leakage of photons (electromagnetic 
radiation quanta) from the MT interior of volume $V$ (`cavity'). 
This leakage may occur from the {\it nodes} of the MT network,
if one assumes fairly isolated interia. 
This leakage will cause decoherence of the coherent state 
of the `MT dimer-OWC' system.
The leakage determines 
the damping time scale $T_r$ in (\ref{decoh}). 

The dimers with their two conformational states
$\alpha$,$\beta$ play the r\^ole of the
collection of $N$ {\it two-level} Rydberg atoms in the 
atomic physics analogue described in section 3.  
If we now make the assumption that the ordered-water 
dipole-quantum coherent modes 
couple to the dimers of the MT chains in a way similar to the 
one leading to a Rabi splitting, described above, then 
one may assume a coupling $\lambda_0$ of order:
\be
   \lambda _0 \sim \frac{d_{dimer} E_{ow}}{\hbar}
\label{mtcoupling}
\ee
where $d_{dimer}$ is the single-dimer electric dipole matrix element,
associated wiith the transition from the $\alpha$ to the $\beta$
conformation, and $E_{ow}$ is a r.m.s. 
typical value of the amplitude of a coherent dipole-quantum field mode. 

Given that each dimer has a mobile 
charge~\cite{hameroff}: $q=18 \times 2e$, $e$ the electron charge, 
one may {\it estimate} 
\be
d_{dimer} \sim  36 \times \frac{\varepsilon_0}{\varepsilon} \times 
1.6. \times 10^{-19} 
\times 4. 10^{-9} \sim 3 \times 10^{-18}~ {\rm Cb} \times {\rm Angstrom} 
\label{dipoledimer}
\ee
where we 
used the fact that a typical distance for the estimate 
of the electric dipole
moment for the `atomic' transition between the $\alpha,\beta$
conformations is of ${\cal O}(4~{\rm nm})$, i.e. of order of the 
distance between the two hydrophobic dimer pockets. 
We also took account of the fact that, as a result of the 
water environment, the electric charge of the dimers appears
to be screened by the relative 
dielectric constant of the water, $\varepsilon/\varepsilon_0 \sim 80$. 
We note, however, that the biological environment of the unpaired 
electric charges in the dimer may lead to 
further suppression of $d_{dimer}$  (\ref{dipoledimer}). 

The amplitude of the collective modes of the dipole-quanta
may be estimated using the formula
(\ref{amplitude}). In our case  the `cavity volume' is:
\be
V \sim 5 \times 10^{-22}~{\rm m}^3 
\label{volumemt}
\ee
which is a typical MT volume, for a moderately long 
MT $L \sim 10^{-6} {\rm m}$, considered as an isolated 
cavity~\footnote{Here we consider only a single unit of the MT network.
The MT network
consists of a large number of such cavities/MT. Our scenario here 
is to examine soliton formation in a single (isolated) 
cavity MT, with the only 
source of dissipation the leakage of photons. 
This would explain the formation of kinks in a single MT~\cite{mtmodel,mn}.
If the entire network
of MT is viewed as a cavity, 
then the volume $V$ is complicated, but in that case the volume 
can be estimated roughly 
as ${\cal N}_{mt}V$, with $V$ the average 
volume of each MT, and ${\cal N}_{mt}$ the number of MT in the 
network population. In such a case, the solitonic state extends over the 
entire network of MT. It is 
difficult to model
such a situation by simple one-dimensional Hamiltonians as in refs. 
\cite{mtmodel,mn}.},
and $\omega_c$ a typical frequency of the dipole quanta collective mode 
dynamics. To estimate 
this frequency we
assume, following the `super-radiance' model of ref. \cite{jibu},  
that the dominant modes are those 
with frequencies in the range 
$\omega_c \sim \epsilon/\hbar $,
where $\epsilon$ is the energy difference between the two principal 
energy eigenstates of the water molecule, which are assumed
to play the dominant r\^ole in the interaction with the 
(quantized) electromagnetic radiation field. 
For the water molecule:
\be
 \hbar \omega_c \sim 4~{\rm meV}
\label{energy}
\ee  
which yields 
\be
     \omega_c \sim \epsilon/\hbar \sim 6 \times 10^{12} s^{-1} 
\label{frequency}
\ee
This is of the same order 
as
the characteristic frequency of the dimers
(\ref{frequency2}), implying that  
the dominant cavity mode and the dimer system are almost in resonance.
Note that this is a feature shared by  
the Atomic Physics systems in Cavities examined in section 3, and thus 
we may apply the pertinent formalism to our system.

From (\ref{amplitude})
one obtains 
for the r.m.s 
$E_{ow}$ in order of magnitude: 
\be
    E_{ow} \sim 10^{4}~{\rm V/m}
\label{eowmt}
\ee 
where we took into account the relative dielectric constant of water 
$\varepsilon/\varepsilon_0 \sim 80$. 

It has not escaped our attention that 
the electric fields of such order of magnitude 
can be provided by the electromagnetic interactions 
of the MT dimer chains, the latter 
viewed as giant electric dipoles~\cite{mtmodel}. 
This may be seen to suggest that the super-radiance 
coherent modes $\omega_c$, which in our scenario 
interact with the unpaired electric charges of the dimers 
and produce the kink solitons along the chains, 
owe their existence to  the 
(quantized) electromagnetic interactions 
of the dimers themselves. 

We assume that the system of ${\cal N}$  MT dimers 
interacts with a {\it single} dipole-quantum mode of the ordered 
water and we ignored interactions among the dimer 
spins~\footnote{More complicated situations, including interactions 
among the dimers, as well as of the dimers with more than 
one radiation quanta,
which might undoubtedly occur
in nature, complicate the above estimate.}.  
In our work here we concentrate our attention 
on the formation of a coherent soliton along 
a single dimer chain, the interactions 
of the remaining 12 chains in a protofilament 
MT cylinder being represented 
by appropriate interaction terms in the effective 
potential of the chain MT model of ref. \cite{mtmodel}.  
In a moderately long microtubule of length $L \simeq 10^{-6}~m$ 
there are 
\be
{\cal N} = L/8 \simeq 10^2
\label{nodimers}
\ee
tubulin 
dimers of average length $8~nm$ in each chain.

Then, from 
(\ref{mtcoupling}), 
(\ref{dipoledimer}), and (\ref{eowmt}), 
the conjectured Rabi-vacuum splitting, describing the entanglement
of the interior coherent modes with the dimers can be estimated to be 
of order 
\be 
    {\rm Rabi~coupling~for~MT} \equiv \lambda _{MT} 
= \sqrt{{\cal N}} \lambda_0 \sim 3 \times 10^{11} s^{-1} 
\label{rabiMT}
\ee
which is, on average, an order of magnitude 
smaller than the characteristic frequency 
of the dimers (\ref{frequency2}). 
In this way, the perturbative 
analysis of section 3, for small  Rabi splittings $\lambda << \omega_0$, 
is valid. Indeed, as can be seen from (\ref{frequency2}),(\ref{frequency}),
the detuning $\Delta$ is of order:
\be 
       \Delta / \lambda_0 \sim {\cal O}(10)-{\cal O}(100) 
\label{detun}
\ee
implying that the condition $\lambda_0^2 {\cal N}/|\Delta|^2 << 1$,
necessary for the perturbative analysis of section 3, is satisfied. 
The Rabi frequency (\ref{rabiMT})  
corresponds to an energy splitting $\hbar \lambda_{MT} 
\sim 0.1~{\rm meV}$ for a moderately long MT.

Having estimated the Rabi coupling between the dimers and the ordered-water
coherent modes
we now proceed to 
estimate the average time scale necessary for the 
formation 
of the pointer coherent states which arise 
due to this entanglement~\cite{zurek,mn}. 
This is the same as the decoherence time 
due to the water-dimer coupling.  
In such a case one 
may use the Liouville model of MT~\cite{mn}, 
discussed in section 2.2, which is known to possess
pointer coherent states~\cite{aspects}.
The relevant decoherence time scale is 
given by (\ref{liouvdecoh}). 

For an accurate estimate of the decoherence time
one should have a precise knowledge
of the 
violations of conformal invariance associated with the 
water-dimer coupling. Unlike the quantum-gravity induced decoherence
case, where a microscopic model for the description of the 
distortions caused in the environment of the chain 
by the external stimuli is available~\cite{mn}, 
at present a precise conformal-field-theory 
model for the description of the water-dimer coupling 
is 
lacking.

However, for our purposes here, 
it suffices to use our 
generic approach to the string-theory
representation of MT chains, described in ref. 
\cite{mn}, where a good order-of-magnitude estimate for $\beta^u$
is provided by (\ref{betau}). We, therefore, merely need to 
have an estimate of the scales $E$ and $M_s$ for the problem at hand.

To estimate the `string' scale $M_s$ in our case~\footnote{In 
ref. \cite{mn}, where scenaria for gravity-induced 
conscious perception were discussed, the scale 
$M_s$ was taken to be the Planck scale $10^{19}$ GeV.
In that case 
decoherence was induced by 
the coupling of the entire network 
of brain MT  to quantum-gravity fluctuations
due to the distortion of the surrounding space time, as result
of abrupt conformational changes in the dimers, caused by external
stimuli~\cite{ideas,HP,mn}.},
where only quantum-electromagnetic interactions enter,  
we notice that in the standard string theory 
$M_s$ acts as an ultraviolet cut-off in the energy
of the effective target-sace field theory, in our case the 
field theory of the 
displacement field $u(x,t)$. 
Since in the ferroelectric chain the closest distance 
separating two unpaired electrons, which are assumed to have the dominant
interactions with the water,
is the size of a tubulin dimer 
$\sim d_{min} = 4~{\rm nm}$,  
an order of magnitiude estimate of $M_s$ is:
\be 
M_s \sim \hbar v_0/d_{min} \sim 
1.5 \times 10^{-4} {\rm eV}, 
\label{Ms}
\ee
where we took into account that 
the Liouville model for MT in ref. \cite{mn} 
is formally  a relativistic string model, but 
with the r\^ole of the velocity of 
light played by the sound velocity $v_0=1 {\rm Km}/{\rm sec}$
in organic biological materials.
It is interesting 
to note that this cut-off in energies 
is less than the kinetic
energies of fast kinks,
propagating with the valocity of sound. Indeed such energies are 
of order $E_{0} \sim 10^{-2} {\rm eV}$. 
Such fast kinks have been associated in ref. \cite{mn}
with quantum gravitational effects, due to abrupt distortions
of space time, 
and hence their exclusion 
in the approach of the present work, by considering kinks
with energies much less than those, as implied by $M_s$, 
is consistent with our 
considering only electromagnetic interactions
among the dimers and its environment.

The above estimate allows us to 
get an idea of the typical energy scales 
of the excitations of the MT systems
that dominate the ordered-water-dimer 
coupling, in the above scenario, and lead to 
decoherence. 
As discussed in ref. \cite{mn}, 
the dominant part of the energy of a 
kink in the model of ref. \cite{mtmodel} 
is of order $1 {\rm eV}$, and thus much higher than 
the $M_s$ (\ref{Ms}). 
Such scales may play a r\^ole 
in the decoherence due to quantum-gravity entanglements~\cite{mn},
which however is much weaker than the electromagnetic ones considered 
here. On the other hand, as we shall argue now, 
a typical energy scale for the dimer displacement field $u(x,t)$, 
much smaller than $M_s$, and therefore 
consistent with our low-energy approach, 
is provided by the kinetic energy of the kink, which is estimated to be 
of order~\cite{mtmodel}
\be
E_{kin} \simeq 5 \times 10^{-8} {\rm eV}
\label{ekin}
\ee
As we shall argue below, 
this energy scale may be used for our estimate of the 
time scale (\ref{liouvdecoh}), 
which is necessary for the formation of the coherent state
due to the water-dimer coupling. 
We stresss that (\ref{ekin}) should not be considered as 
a typical energy scale associated with the excitation 
spectrum of the dimers, pertaining to the Rabi splitting
discussed above. 
It is rather an `effective scale', characterising  `friction' effects 
between the dimers and the water environment. The latter are
responsible~\cite{mn} for 
decoherence and the eventual formation 
of the kink coherent states, which are 
minimum-entropy states~\cite{zurek}, 
least susceptible to the effects of the water environent. 

To justify the above estimate, one should notice that 
the formation of coherent quantum 
states through decoherence due to friction~\cite{zurek} 
is the quantum analogue of 
the `drift velocity' acquired by a Brownian particle 
in classical mechanics.
In the (non-relativistic) conformal-field-theory setting of ref. \cite{mn}
such a friction could be described by the formation of point-like defects
on the dimer chains, which could be described by appropriate non-relativistic 
membrane backgrounds in the $1 + 1$-dimensional string theory representation 
of the MT dynamics. Such membranes are stringy defects, which, 
for instance, could 
describe the result of 
an abrupt conformational change of a given dimer due to its coupling 
with the water environment. Scattering of stringy excitations $u(x,t)$
off the defect causes `recoil' of the latter, which starts moving with a 
velocity $v_d$. The recoil is due to quantum fluctuations as argued in ref. 
\cite{emnrecoil}. Such recoil effects are necessary for energy and momentum 
conservation 
in the case of (non-relativistic) heavy membranes,
considered in ref. \cite{emnrecoil}.
That model 
represents a pilot model for discussing decoherence due to water 
environment in our case. 

The important feature to notice is that the `effective mass' of the defect 
in such a stringy representation turns out to be inverssely proprotional 
to the string-coupling constant $g_s$~\cite{emnrecoil}, 
\be
       m_{defect} =(8\sqrt{2}\pi g_s)^{-1}
\label{effmass}
\ee
which, in turn, depends on the vacuum expectation 
value of the effective `dilaton' field $\Phi$: $g_s=e^{-<\Phi>}$. 
In the case at hand, as discussed in ref. \cite{mn}, 
the dilaton field is proprotional to the friction coefficient $\gamma$, 
so that 
$g_s \sim {\rm exp}(-\gamma)$ in order of magnitude. Strong friction,
therefore, between water molecules and dimers, 
which may be assumed in our physical model of MT, 
implies weak string coupling, and hence the non-relativistic approximation 
for the membranes proves sufficient. 

Within the framework
of identifying target time with the Liouville mode~\cite{emn}, then,  
the analysis of \cite{emnrecoil} has shown that the {\it quantum recoil}
degrees of freedom of the membrane defect carry {\it information} and induce decoherence
of the dimer (string) subsystem, corresponding to violations of conformal 
invariance of order: 
\be
    \beta^{recoil} \sim {\cal O}(v_d^2/16\pi g_sM_s)
\label{recoilveloc}
\ee
where $v_d^2$ is the recoil velocity of the defect. By energy-momentum conservation, 
the kinetic energy of the recoiling (non-relativistic) defect
is {\it at most} of the same order as the kinetic energy of the 
displacement field (\ref{ekin}). 
This would yield a lower bound on the 
decoherence time (\ref{liouvdecoh}) estimated on the basis 
of (\ref{recoilveloc}) to be of order:
\be 
t_{owdecoh} \gsim   10^{-10}  {\rm sec} 
\label{Mhz}
\ee
for a moderately long MT, with ${\cal N}=10^2$ dimers.
This is the time scale over which 
solitonic coherent pointer states in the MT 
dimer system are formed (`pumped'), according to our scenario. 
Note that the 
scale (\ref{Mhz}) 
is not far from the original Fr\"ohlich scale~\cite{Frohlich}, 
$10^{-11}-10^{-12}~{\rm sec}$.

To  
answer the question whether quantum coherent pointer states 
are responsible for loss-free energy transport 
across the MT
one should 
examine the time scale of the decoherence induced by 
the coupling of the MT to their biological environment 
as a consequence of 
dissipation through the walls of the MT cylinders.
Such an ordinary environmental entanglement
has been ignored in the derivation of (\ref{Mhz}).
It is this environment that will induce decoherence 
and eventual collapse of the pointer states formed
by the interaction of the dimers with the coherent modes
in the ordered water. 
Using typical numbers of MT networks, we can estimate
this decoherence time 
in a way similar to the corresponding
situation in atomic physics (\ref{decoh},\ref{distance}): 
\be
    t_{collapse} = \frac{T_r}{2 n {\cal N}{\rm sin}^2\left(\frac{{\cal N}
n \lambda_0^2t}{\Delta}\right)} 
\label{convtime}
\ee
where we took into account 
that 
the dominant (dimer)-(dipole quanta) coupling 
occurs for ordered-water `cavity' modes which are {\it almost at resonance}
with the dimer oscillators (c.f. (\ref{frequency}),(\ref{frequency2})),
slightly detuned by $\Delta:~\lambda_0/\Delta 
<<1 $, c.f. (\ref{detun}).

We also assume that a typical coherent mode of dipole/quanta
contains an average of $n = {\cal O}(1)-{\cal O}(10)$  
oscillator quanta. 
The {\it macroscopic} character of the Schr\"odinger's cat
dimer-dipole-quanta system comes from the {\cal N} dimers 
in a MT (or ${\cal N}_{mt}{\cal N}$ in MT networks)~\cite{mn}. 

The time $t$ appearing in (\ref{convtime})
represents the `time' of interaction of the dimer system with the 
dipole quanta. A reasonable estimate of this time scale 
in our MT case can be obtained by 
equating it with the average 
life-time of 
a coherent dipole-quantum state, which, in the 
super-radiance model of ref. \cite{jibu}
can be estimated as
\be
     t \sim \frac{c\hbar ^2V}{4\pi d_{ej}^2\epsilon N_{w}L}
\label{lifetime}
\ee
with $d_{ej}$ the electric dipole moment of a water molecule, 
$L$ the length of the MT, and $N_w$ the number of water molecules 
in the volume $V$ of the MT. For typical values of the parameters
for moderately long MT, $L \sim 10^{-6}~{\rm m}$, $N_w \sim 10^8$, 
a typical 
value of $t$ is: 
\be 
         t \sim 10^{-4}~{\rm sec} 
\label{lifetime2}
\ee

We remark at this point that this 
is considerably larger than 
the average life-time of a coherent dipole quantum 
state 
in the water model 
of ref. \cite{prep}. Indeed in that model, 
the corresponding life-time is estimated to be:
\be 
     t \sim 2\pi/\omega_0 
\label{waterdecoh}
\ee
where $\omega_0 \sim 1/I$, is a typical frequency 
of resonating electromagnetic mode in the ordered water.
For a typical value of the water molecule moment of intertia~\cite{prep}
this yields a time-scale associated with the coherent
interaction ${\cal O}(10^{-14}~{\rm sec})$.

The time scale $T_r$, over which a cavity MT dissipates its energy,  
can be identified in our model with the average life-time 
(\ref{lifetime}) of a coherent-dipole quantum state: 
\be
T_r \sim t \sim  10^{-4}~{\rm sec}
\label{trfrohlich}
\ee
which leads to a naive estimate 
of the quality factor for the MT cavities,
$Q_{MT} \sim \omega_c T_r \sim {\cal O}(10^8)$. 
We note, for comparison, that high-quality 
cavities encountered in Rydberg atom
experiments 
dissipate energy in time scales of ${\cal O}(10^{-3})-{\cal O}(10^{-4})$
sec, and have $Q$'s which are comparable to $Q_{MT}$ above. 
Thus, it is not unreasonable to expect that conditions 
like (\ref{trfrohlich}),
characterizing MT cavities, are met in Nature.

{}From (\ref{convtime}), (\ref{rabiMT}), and (\ref{trfrohlich}), 
one then obtains the following estimate 
for the collapse time of the kink coherent state of the MT dimers 
due to dissipation:
\be
t_{collapse} \sim {\cal O}(10^{-7})-{\cal O}(10^{-6})~{\rm sec}
\label{tdecohsoliton}
\ee
which is larger or equal than the Fr\"ohlich scale (\ref{FS})
required for energy transport 
across the MT by an average kink soliton in the model of ref. 
\cite{mtmodel}. The result (\ref{tdecohsoliton}), then, 
implies that 
Quantum Physics may be responsible for 
dissipationless energy transfer across the MT. 

We close this section by remarking that 
if the condition $T_{collapse} \gsim 5 \times 10^{-7} {\rm sec}$
is not met, then 
the above picture, based on mesoscopic coherent states, 
would be inconsistent 
with energy-loss free energy transport, since 
decoherence due to environmental entanglement 
would occur before energy could be transported across the 
MT by the preformed quantum soliton.
In such a case energy would be transported 
due to different mechanisms, one of which is 
the 
classical solitons scenario of ref. \cite{mtmodel,lal}. 
However, even in such cases of fast decoherence, the 
Rabi coupling predicted above could still exist and be 
subjected to experimental verification.

\section{Outlook: Possible Experimental Tests Using 
Micro-Tubules as Cavities}

In the present work we have put forward a conjecture
concerning the representation of the MT arrangements 
inside the cell as {\it isolated} high-Q(uality) {\it cavities}. 
We presented a scenario according to which 
the presence of the ordered water in the interior of the cylindrical 
arrangements results in the appearance of electric dipole quantum 
coherent modes, which couple to the unpaired electrons of the 
MT dimers via Rabi vacuum field couplings, familiar from 
the physics of Rydberg atoms in electromagnetic cavities~\cite{rabi}. 
In quantum optics, such couplings are considered as experimental 
proof of the quantized nature of the electromagnetic radiation.
In our case, therefore, if present, such couplings could 
indicate the existence of the coherent quantum modes
of electric dipole quanta in the ordered water environment of MT,
conjectured in ref. \cite{delgiud,prep}, and used in the present work.

Some generic decoherence time estimates, due to environmental 
entanglement of the MT cavities, have been given. The conclusion was 
that only in {\it fairly isolated} cavities, which we conjecture exist
inside the biological cells, 
decoherence occurs in time scales which are in agreement 
with the Fr\"ohlich scale ($\sim 5 \times 10^{-7}$ sec) for 
energy transfer across the MT via the formation 
of kink solitonic structures. In such a case,  
{\it dissipationless energy transfer} might occur in biological systems,
an in particular in MT which are the substratum of cells, 
in much the same way as frictionless electric current transport occurs
in superconductors, i.e. via {\it quantum coherent modes} that extend over 
relatively large spatial regions. A phenomenological analysis 
indicated that for moderately long MT networks such a situation could be 
met if the MT cavities dissipate energy in time scales 
of order $T_r \sim 10^{-4}-10^{-5} {\rm sec}$. This lower bound 
is comparable to the 
corresponding scales of atomic cavities~\cite{rabiexp}, 
$\sim 10^{-4} {\rm sec}$, which, in turn, implies that the 
above scenaria, on dissipationless energy transport 
as a result of the formation of quantum coherent states,
have a good chance of being realized at the scales 
of MT, which are comparable to those in Atomic Physics.

We have conjectured in this work
that an indirect verification of such a mechanism would be the 
experimental detection of the aforementioned Vacuum field Rabi coupling,
$\lambda_{MT}$,  
between the MT dimers and the ordered water quantum coherent modes. 
This coupling, if present,  
could be tested 
experimentally  
by the same methods used to measure VFRS in atomic physics~\cite{rabiexp}, 
i.e. by using the MT themsleves as {\it cavity environments}, 
and considering tuneable probes to excite the coupled dimer-water
system. Such probes could be pulses of (monochromatic ) light, for example,
passing through the hollow cylinders of the MT. 
This would be the analogue of an external field in the atomic experiments
described above, which would then resonate, not at the bare frequencies 
of the coherent dipole quanta or dimers, but at the {\it Rabi splitted} ones,
and this would have been exhibited by a double pick in the absorption 
spectra of the dimers~\cite{rabiexp}. By using MT of different sizes
one could thus check on 
the characteristic $\sqrt{N}$-enhancement of the Rabi coupling 
for MT systems with $N$ dimers. 

The technical complications 
that might arise in such experiments are associated 
with the absence of completely
resonant cavities in practice. In fact, from our discussion in this article,  
one should 
expect a slight detuning $\Delta$ between the cavity mode and 
the dimer, of  
frequency $\omega _0$. As discussed in section 3,
the detuning produces a split of the vacuum-Rabi 
doublet into a cavity line $\omega _0+ \lambda^2/\Delta $ and 
an `atomic line' $\omega _0- \lambda^2/\Delta $.  
In atomic physics there are well established 
experiments~\cite{rabiexp,rabiexp2} to detect such splittings.
In fact, detection of such lines is considered as 
a very efficient way of `quantum non-demolition' measurement~\cite{qnd}
for small microwave photon numbers. Such atomic physics experiments, 
therefore,
may be used as a guide in 
performing the corresponding biological experiments 
involving MT (as cavities), as suggested in the present work.

We believe that 
such experimental set-ups will help in 
sheding some light on the issue of 
the quantum mechanical nature of the MT arrangements, and even 
on 
the processes of the transmission of electric signals (stimuli) 
by the neuronic systems. 
Clearly much more work needs to be done before even tentative
conclusions are reached, concerning the nature of the MT 
arrangements. However, we believe that the present work 
constitutes a useful addition to the programme of understanding 
the nature of the MT arrangements inside the cell, and the associated
processes of energy transfer across the cells. 

\pr
\nk {\Large {\bf Acknowledgements}}
\pr
The work of D.V.N. is supported
in part by D.O.E. Grant
DEFG05-91-GR-40633.


\begin{thebibliography}{99}

\bibitem{hameroff} P. Dustin, {\it MicroTubules}
(Springer, Berlin 1984);
\par Y. Engleborghs, {\it Nanobiology} 1 (1992), 97.

\bibitem{Frohlich} H. Fr\"ohlich, {\it Bioelectrochemistry},
ed. by F. Guttman and H. Keyzer (Plenum, New York 1986).


\bibitem{lal} P. Lal, Physics  Letters 111A (1985), 389.


\bibitem{mtmodel} M.V. Satari\'c, J.A. Tuszy\'nski,
R.B. Zakula, Phys. Rev.  E48 (1993), 589;
\par this model of MT dynamics 
is based on the ferroelectric-ferrodistortive model 
of : M.A. Collins, A. Blumen,
J.F. Currie, and J. Ross, Phys. Rev. B19 (1978), 3630..

\bibitem{mn} N.E. Mavromatos and D.V. Nanopoulos, 
Int. J. Mod. Phys. B11 (1997), 851. 


\bibitem{zurek} For a comprehensive review see: W.H. Zurek, 
Phys. Today 44, No. 10 (1991), 36; 
see also: W.H. Zurek, Phys. Rev. D24 (1981), 1515;
\par A.O. Caldeira and A. J. Leggett, Physica (Amsterdam)
121A (1983), 587; Ann. Phys. 149 (1983), 374.  


\bibitem{emn} J. Ellis, N.E. Mavromatos
and D.V. Nanopoulos, Phys. Lett. B293 (1992), 37;
\par {\it lectures presented at the
Erice Summer School, 31st Course: From Supersymmetry to the
Origin of Space-Time},
Ettore Majorana Centre, Erice, July 4-12
1993 ; hep-th/9403133, 
Vol. 31 (1994), p.1 
(World Sci. );
\par For a pedagogical review of this approach see:
D.V. Nanopoulos, Riv. Nuov. Cim. Vol. 17, No. 10 (1994), 1.




\bibitem{HP} S. Hameroff and R. Penrose, 
{\it Orchestrated Reduction of Quantum Coherence
in Brain Microtubules: a Model of Consciousness},
in {\it Towards a science of Consciousness}, The First Tucson
Discussions and Debates, eds. S. Hameroff {\it et al.} (MIT Press,
Cambridge MA 1996), p. 507-540.



\bibitem{ideas} R. Penrose, {\it The Emperor's New Mind}
(Oxford Univ. Press 1989); {\it Shadows of the Mind}
(Oxford Univ. Press 1994);
\par D.V. Nanopoulos, {\it Theory of Brain
Function, Quantum Mechanincs and Superstrings},
hep-ph/9505374,
Proc. ``XV Brazilian National Meeting
on Particles and Fields'', eds. M.S. Alves {\it et al.}, 
Brasilean Phys. Society (1995), p. 28. 

\bibitem{ehns} J. Ellis, J.S. Hagelin, D.V. Nanopoulos and
M. Srednicki, Nucl. Phys. B241 (1984), 381.

\bibitem{emohn} J. Ellis, S. Mohanty and D.V. Nanopoulos,
Phys. Lett. B221 (1989), 113.



\bibitem{gisin} N. Gisin and I. Percival, J. Phys. A26 (1993), 2233.

\bibitem{albrecht} A. Albrecht, Phys. Rev. D46 (1992), 5504.

\bibitem{aspects} J. Ellis, N.E. Mavromatos and 
D.V. Nanopoulos, Proc. {\it 1st International Workshop 
on Phenomenology of Unification from Present to Future},
23-26 March 1994, Roma (eds. G. Diambrini-Palazzi {\it et al.},
World Sci., Singapore 1994), p.187. 


\bibitem{guides} S.R. Hameroff, Am. J. Clin. Med. 2 (1974), 163. 

\bibitem{ordered} S. Hameroff, {\it Ultimate Computing}
(Elsevier North-Holland,
Amsterdam 1987);
\par S. Hameroff, S. A. Smith, R.C. Watt,
Ann. N.Y. Acad. Sci. 466 (1986), 949.


\bibitem{prep} E. Del Giudice, G. Preparata and G. Vitiello,
Phys. Rev. Lett. 61 (1988), 1085.


\bibitem{delgiud} E. Del Giudice, S. Doglia, M. Milani
and G. Vitiello, Nucl. Phys. B251 (FS 13) (1985), 375;
{\it ibid} B275 (FS 17) (1986), 185.

\bibitem{jibu} M. Jibu, 
S. Hagan, S. Hameroff, K. Pribram and 
K. Yasue, Biosystems 32 (1994), 195. 

\bibitem{rabi} J.J. Sanchez-Mondragon, N.B. Narozhny 
and J.H. Eberly, Phys. Rev. Lett. 51 (1983), 550.



\bibitem{rabiexp} F. Bernardot {\it et al.}, Europhysics Lett. 
17 (1992), 34.


\bibitem{rabiexp2} M. Brune {\it et al.}, Phys. Rev. Lett. 65 (1990), 976. 

\bibitem{otinowski} M. Otwinowski, R. Paul and 
W.G. Laidlaw, Phys. Lett. A128 (1988), 483. 


\bibitem{DDK}F. David, Mod. Phys. Lett. A3 (1988), 1651;
\par J. Distler and H. Kawai, Nucl. Phys. B321 (1989), 509.


\bibitem{aben} I. Antoniadis, C. Bachas, J. Ellis
and D.V. Nanopoulos, Phys. Lett. B211 (1988), 393;
Nucl. Phys. B328 (1989), 117; Phys. Lett. B257 (1991), 278;
\par See also D.V. Nanopoulos, in {\it Proc. International
School of Astroparticle Physics}, HARC (Houston) (World
Scientific, Singapore, 1991), p. 183.

\bibitem{haroche} S. Haroche and J.M. Raimond, 
{\it Cavity Quantum Electrodynamics}, 
ed. P. Berman (Academic Press, New York 1994), p.123,
and references therein.  


\bibitem{agar} G.S. Agarwal, Phys. Rev. Lett. 
53 (1984), 1732. 


\bibitem{dual} Yifu Zhu {\it et al.}, Phys. Rev. Lett. 64 (1990), 2499. 

\bibitem{brune} M. Brune {\it et al.}, Phys. Rev. Lett. 
77 (1996), 4887.

\bibitem{qnd} M. Brune {\it et al.}, Phys. Rev. A45 (1991), 5193.

\bibitem{markov} Y.R. Shen, Phys. Rev. 155 (1967), 921;
\par A.S. Davydov and A.A. Serikov, Phys. Stat. Sol.
B51 (1972), 57;
\par B. Ya. Zel'dovich, A.M. Perelomov and V.S. Popov, 
Sov. Phys. JETP 28 (1969), 308;
\par For a comprehensive review see: V. Gorini {\it et al.},
Rep. Math. Phys. Vol. 13 (1978), 149. 

\bibitem{milburn} D.F. Walls and G.J. Milburn, Phys. Rev. A31 (1985), 2403.


\bibitem{emnrecoil} J. Ellis, N.E. Mavromatos and D.V. Nanopoulos, 
hep-th/9609238; and
hep-th/9704169, Mod. Phys. Lett. A, in press. 


\end{thebibliography}
\end{document}